\documentclass[final,12 pt, twocolumn]{elsarticle}
\usepackage{soul}
\usepackage{natbib}
\usepackage{lineno}
\usepackage{cite}
\usepackage{hyperref}       
\usepackage{url}
\usepackage{booktabs}       
\usepackage{amsfonts}    
\usepackage{amssymb}
\usepackage{amsmath}
\usepackage{graphicx}
\usepackage{caption}
\usepackage{algpseudocode, algorithmicx, algorithm,mismath}
\usepackage{siunitx}
\usepackage{multirow}
\usepackage{pdfpages}
\usepackage{svg}
\usepackage{subfig}
\usepackage{changes}
\renewcommand\cite{\citep}

\newcommand \andothers {\textit{et~al.}}
\newcommand \subseq {W}
\newcommand \Fig {Fig.}
\newcommand{\mathbold}[1] {\textbf{\textit{#1}}}
\modulolinenumbers[5]
\journal{Energy}

\usepackage{url}

\begin{document}
\begin{frontmatter}
\title{Electric Vehicle Identification from Behind Smart Meter Data}

\author[1]{\href{https://orcid.org/0000-0002-7441-9344}{\includegraphics[scale=0.06]{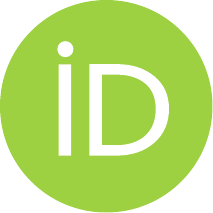}\hspace{1mm} Ammar Kamoona\corref{cor1}}}
\ead{ammar.kamoona@rmit.edu.au}
\cortext[cor1]{Corresponding author}
\author[1]{Hui Song}
\author[1]{Ali Moradi Amani}
\author[1]{~\href{https://orcid.org/0000-0002-0517-9420}{\includegraphics[scale=0.06]{orcid.pdf}\hspace{1mm} Mahdi Jalili}}
\author[1]{Xinghuo Yu}
\author[2]{Peter McTaggart} 
\affiliation[1]{organization={School of Engineering, RMIT University},
postcode={3000 VIC},
city={Melbourne},
country={Australia}}
\affiliation[2]{organization={CitiPower and Powercor},
city={Melbourne},
country={Australia}}
\fntext[fn1]{Ammar Kamoona is currently working as a Postdoctoral researcher, RMIT University, Victoria, Australia.}
\fntext[fn2]{This work is supported by Australian Research Council through
 project NO. LP180101309 and Victorian Government through Victorian
 Higher Education State Investment Fund (Supporting electrification of
Victoria’s future fleet)}

\newpageafter{author}
\begin{abstract}
Electric vehicle (EV) charging loads identification from behind smart meter recordings is an indispensable aspect that enables effective decision-making for energy distributors to reach an informed and intelligent decision about the power grid's reliability. When EV charging happens behind the meter (BTM), the charging occurs on the customer side of the meter, which measures the overall electricity consumption. In other words, the charging of the EV is considered part of the customer's load and not separately measured by the Distribution Network Operators (DNOs). DNOs require complete knowledge about the EV presence in their network. Identifying the EV charging demand is essential to better plan and manage the distribution grid. Unlike supervised methods, this paper addresses the problem of EV charging load identification in a non-nonintrusive manner from low-frequency smart meter using an unsupervised learning approach based on anomaly detection technique. Our approach does not require prior knowledge of EV charging profiles. It only requires real power consumption data of non-EV users, which are abundant in practice. We propose a deep temporal convolution encoding decoding (TAE) network. The TAE is applied to power consumption from smart BTM from Victorian households in Australia, and the TAE shows superior performance in identifying households with EVs.    
\end{abstract}

\end{frontmatter}


\section{Introduction}
%
%
%
%

Transport electrification is one of the major strategies in achieving zero-net emission targets. The transport sector alone is responsible for about one-quarter of energy-related emissions~\cite{yong2015review}. To decarbonize the transport sector, Electric Vehicle (EV) adoption has increased in the market in recent years. The Energy Outlook predicts that about 100 million EVs will be on the road worldwide by 2035~\cite{noauthor_prospects_nodate}. Governments worldwide are taking different programs, initiatives, and policies to increase EVs uptake.  

The increase in EV uptake has caused technical challenges in the power distribution grids in terms of power demand increase, system losses, voltage drops, phase unbalance, and stability issues~\cite{green2011impact,yong2015review}. To cope with the impact of the large-scale EV penetration on the  power distribution grids, different approaches have been conducted, mainly focusing on demand-response and public charging stations. These approaches range from EV coordinated charging using optimal charging price~\cite{zhang2018optimal}, or power allocation optimization~\cite{YanEV2021}, and EV charging location planning for competitive resource allocation~\cite{li2021data}. Most of these works are based on the assumption that EV charging is mainly done at public stations. However, many EV owners charge their EVs at home due to convenience and flexibility in choosing the charging time and often lower charging costs during off-peak hours. A recent Australian study showed that even 47\% of the business fleets are usually home garaged, and thus home charging facilities are required for them~\cite{mortimore2022business}.

Unmanaged home charging can greatly impact the power grid during peak hours, causing low power quality in voltage~\cite{li2012impacts}. DNOs and retailers are interested in overcoming the impact using smart charging and scheduling algorithms~\cite{lee2021adaptive}. However, EV identification, i.e., detecting EV in a household using its smart meter data, is the main prerequisite. This data can help DNOs to understand the impact of EVs on the local grid and to identify opportunities for grid optimization. For example, behind-the-meter data can help DNOs to identify households with excess energy storage capacity in their EV batteries, which can be used to provide grid services such as frequency regulation or peak shaving. Additionally, behind-the-meter data can help DNOs to understand the charging behavior of EV owners and to develop targeted interventions to encourage off-peak charging or to manage peak demand. Overall, knowing which households have EVs and having access to behind-the-meter data can help DNOs to manage the impact of EVs on the grid and to develop more efficient and sustainable energy systems. 
There are limited research articles in the direction of EV identification from behind smart meter data that do not require any knowledge about the charging events profile, arrival time, departure time, and power demand. The latest research paper~\cite{HuiEV2021iit} on EV identification has tackled this problem using supervised learning by building an ensemble classifier using non-EV and EV users. Due to the limited availability of EV users and to avoid the problem of data imbalance, Hui~\cite{HuiEV2021iit}~\andothers~proposed a clustering approach for non-EV user data to generate more representative data (under-sampling) by using pattern recognition or feature extraction. Feature extraction from smart meter data has been used for EV driving pattern recognition~\cite{hu2020intelligent,munshi2018unsupervised,xiang2021charging}, and load profile identification~\cite{yan2020time,xiang2020slope,wang2020deep}. Different features from power consumption data are used. For example, in~\cite{yan2020time}, the average power consumption of weekdays is used. Xiang~\andothers~\cite{xiang2020slope} proposed to use shape features of load profile combined by K-means clustering for load profile identification.

There are significant challenges in working with smart meter data. Information received by DNOs can be incomplete and lag behind. Advanced machine learning techniques to classify \textit{EV} and \textit{non-EV} users may suffer from an imbalance of data due to the slow uptake of EVs in the grid ~\cite{hwang2019hexagan}. Different approaches to balance data can be adopted to address this problem, such as over-sampling~\cite{ye2020oversampling} or under-sampling~\cite{HuiEV2021iit}. Both methods have their own disadvantages. For example, under-sampling excludes the majority of samples, which can potentially reduce model performance. Over-sampling requires a high computation cost, and the redundancy in the minor class may reduce model classification accuracy~\cite{HuiEV2021iit}. Recently, Hamidreza \andothers~\cite{jahangir2021novel} proposed to address the problem of different EV demand characteristics using 3-D convolution via a semi-supervised approach using Generative Adversarial Networks (GANs). 

Unlike supervised learning, unsupervised learning techniques can provide a more suitable and efficient approach that does not require minor class data during the training. Different unsupervised /non-intrusive works have been proposed in the literature~\cite{wang2020deep,munshi2018unsupervised,xiang2021charging} for EV charging profile identification. Wang~\andothers~\cite{wang2020deep} proposed a deep generative model for EV charging profile extraction using Markov processes. The aforementioned paper requires complete knowledge of EVs' arrival and departure times, which DNOs lack access to these data in most cases. Munshi~\andothers~\cite{munshi2018unsupervised} proposed using Independent Component Analysis (ICA) to decompose the smart meter data and extract EV charging load pattern. However, their approach requires different extraction processes to identify different charging patterns. Xiang~\andothers~\cite{xiang2021charging} proposed training free charging load profile extraction by applying two-stage signal decomposition. However, the earlier approaches are based on the assumption that the EV load profile is very distinctive from others. Charging characteristics, such as the EV power consumption and the rectangle profile of charging/discharging, are known. In practice, information received by the energy distributors regarding a new EV charging occurrence can be incomplete, and lagging and charging profiles can not be generalized to all cases when different EV models and with different EV charging stations. In our dataset provided by Victorian DNOs, the same EV model is charged with different charging stations, which results in different charging profiles. More important DNOs cannot access information about charging profiles, arrival time, and departure time of EVs even the testset does not have such information.  Another approach that does not require prior knowledge about the charging profile is anomaly detection, a branch of unsupervised learning.

Anomaly detection, which is the main focus of this paper, is the process of detecting instances in the data that deviate from the predefined normal model~\cite{chandola2009anomaly}. Anomaly detection has been applied to diverse domains such as security, risk management, health, and medical risks~\cite{chandola2009anomaly}. It has also been applied to identify faulty appliances in buildings ~\cite{yip2018anomaly}, electricity theft~\cite{arjunan2015multi,yeckle2018detection}, and meter failure~\cite{yip2018anomaly}. The aforementioned methods aim at finding an EV charging session that is highly based on the assumption that an EV is available in the household. However, in our problem setting, we do not know whether the household/user has an EV, and we want to identify the household with an EV regardless of when the charging session occurred.

The existing anomaly detection in power/energy consumption using smart meter recordings can be roughly classified into the following approaches, clustering-based methods, one-class learning-based methods, and dimension reduction-based methods. Clustering-based methods are based on the assumption that normal data are tightly clustered to one or more clusters, and anomalous data are far from these clusters. Arjunan~\andothers~\cite{arjunan2015multi} proposed a clustering algorithm that comprises two steps to detect an anomaly due to energy theft. Yeckle~\andothers~\cite{yeckle2018detection} proposed power theft detection based on mutual k-nearest neighbour and k-means clustering. One-class classification-based methods are based on building a boundary decision around the normal pattern. Anomaly detection is done based on the distance from the test sample to the decision boundary. One popular example is One-Class-Support Vector Machine (OCSVM). Anomaly detection in a smart home environment using OCSVM is proposed in~\cite{jakkula2011detecting}. Dimension deduction-based methods are based on the assumption that by projecting data from one manifold to a compressed/latent manifold, anomalous samples can not be reconstructed due to the fact that they do not belong to the latent manifold~\cite{sial2021detecting}.

Inspired by the success of temporal convolution~\cite{farha2019ms} in the
sequence modeling, we propose a deep temporal autoencoder (TAE) for EV load charging identification. Different from~\cite{wang2020deep,jahangir2021novel}, our approach is based on identifying the existence of EVs in the households from behind-meter (BTM) data rather than extracting the charging profile. In a BTM scenario, the electricity used to charge the EV is drawn from the customer's electricity supply, which may be from the local distribution network or generated on-site through solar panels or other renewable sources. Therefore, the amount of electricity used to charge the EV is recorded as part of the customer's overall electricity consumption rather than as a separate and distinct metered event. Our dataset lacks information about the power demand, arrival, and departure times of EVs from households, which is also a common problem for most DNOs who aim to identify households with EVs. To the best of our knowledge, no work in the literature has explored the usage of deep anomaly detection for EV charging identification from smart meter data. The contributions of the paper are summarized as follows: 
\begin{itemize}
	\item We propose an unsupervised learning framework for EV charging identification from smart meter data that does not require manual feature engineering or prior knowledge about charging profiles. 
	\item An anomaly-based approach is proposed for EV identification using the deep temporal encoding-decoding network in which nonlinear features representation and learning are done jointly.
	\item We propose using the sum squared of the model $L_2$ loss function to identify charging occurrence in the household. Our experiments show it is more efficient than using the Soft-Dynamic time wrapping loss function.
\end{itemize}

The rest of this paper is organized as follows: Section~\ref{sec:proposed_approach} covers the proposed model for EV charging identification. Section~\ref{Sec:experiments} presents experimental results, and Section~\ref{Sec:conclusion} concludes the paper.

\section{EV identification as an anomaly detection problem}
\label{sec:proposed_approach}
\begin{figure*}
    \centering
    \includegraphics[width=0.9\textwidth]{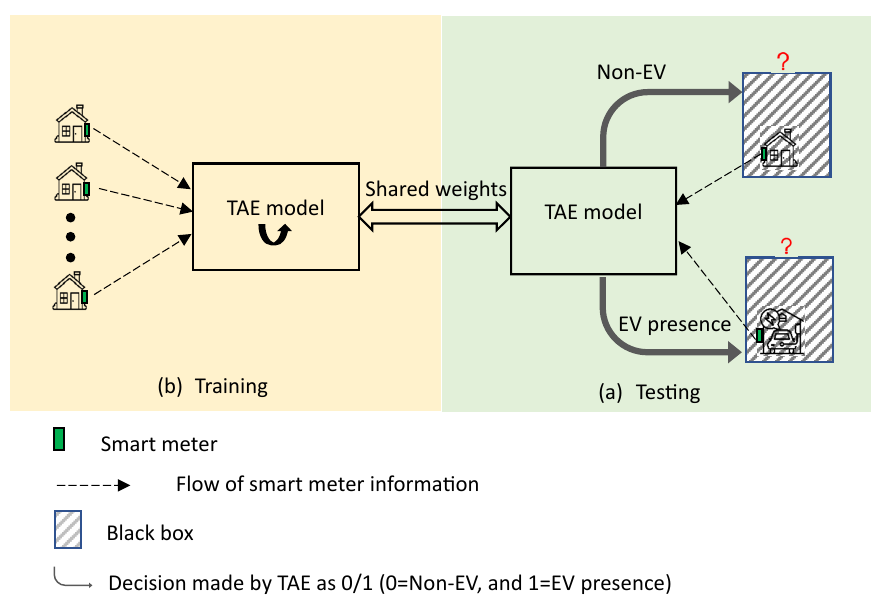}
    \caption{Overview of EV identification using the proposed Temporal Encoder-Decoder (TAE) network. (a) : shows the training phase of TAE where only recording of non-EVs smart meter data is used; (b) testing phase where TAE makes decisions about an unknown household whether it has EV or not.}
    \label{fig:general}
\end{figure*}

In this section, we first introduce the EV identification problem and then propose an anomaly detection approach to solve the problem. The general framework of EV identification is shown in Fig.~\ref{fig:general}. The network consists of the training and testing phases, in which the model weights are shared between both models.

\subsection{EV identification problem}
\label{sec:problem_def}
Let us now formulate the EV identification problem, given $N$ set of consumers' data $X=\{\mathbold{x}_i\}_{i}^{N}$ from smart meter recordings. Each consumer data $\mathbold{x}_i \in \mathbb{R}^{1\times T}$ is a sequence of ordered observations for a period of time, $\mathbold{x}_i=\{x_{i,t},t=1,2,\ldots,T\}$, where $T$ is the length of time sequence. Each consumer's data $\mathbold{x}_i$ is associated with label $y_i\in\{0,1\}$, as a consumer without EV and with EV, respectively. If $y_i=0$, it indicates that none of the recordings have EV charging, whereas $y_i=1$, implies that at least one charging event has occurred. The task here is to build an anomaly detector that assigns a binary label (0 for consumer without EV (normal) and 1 for consumer with EV (abnormal)). In  our anomaly detection approach, the model is trained only using normal samples, i.e., non-EV consumers. 

\subsection{EV charging characteristics}
\begin{figure*}
    
    \subfloat[High demand EV user.]{\includegraphics[width=0.9\textwidth]{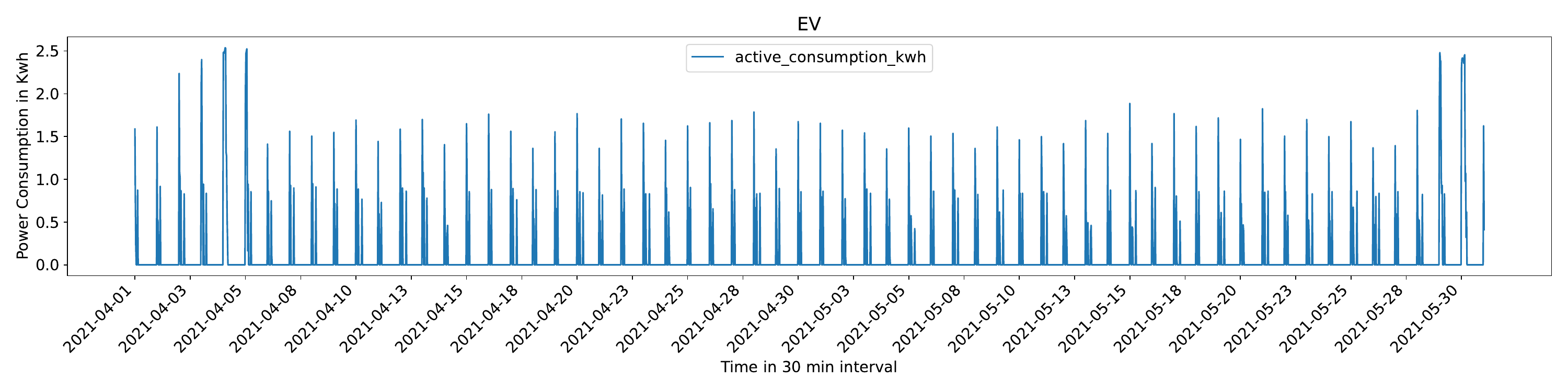}}\\
    \subfloat[Low demand EV user.]{\includegraphics[width=0.9\textwidth]{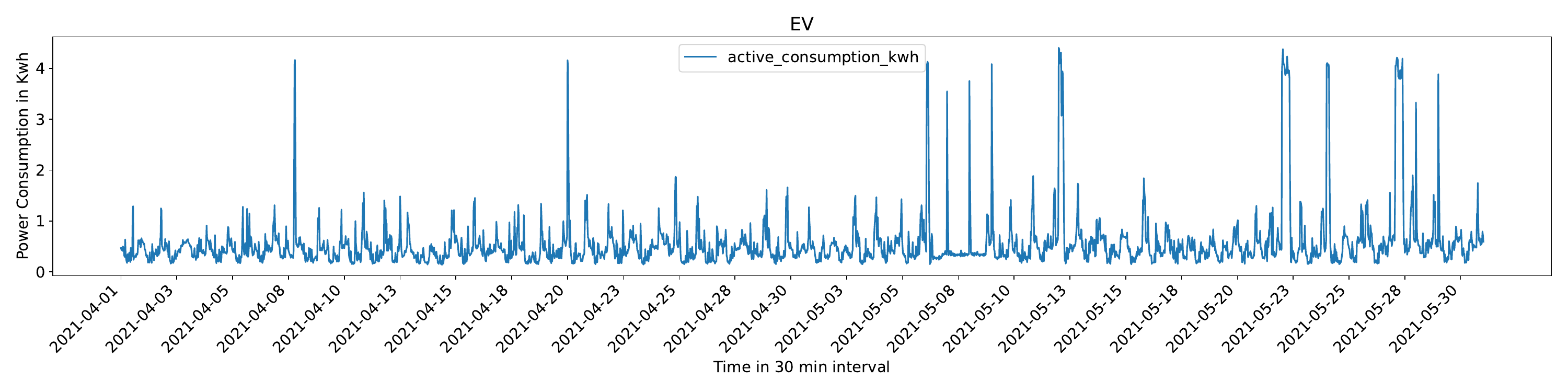}}\\
    \subfloat[Non-EV user]{\includegraphics[width=0.9\textwidth]{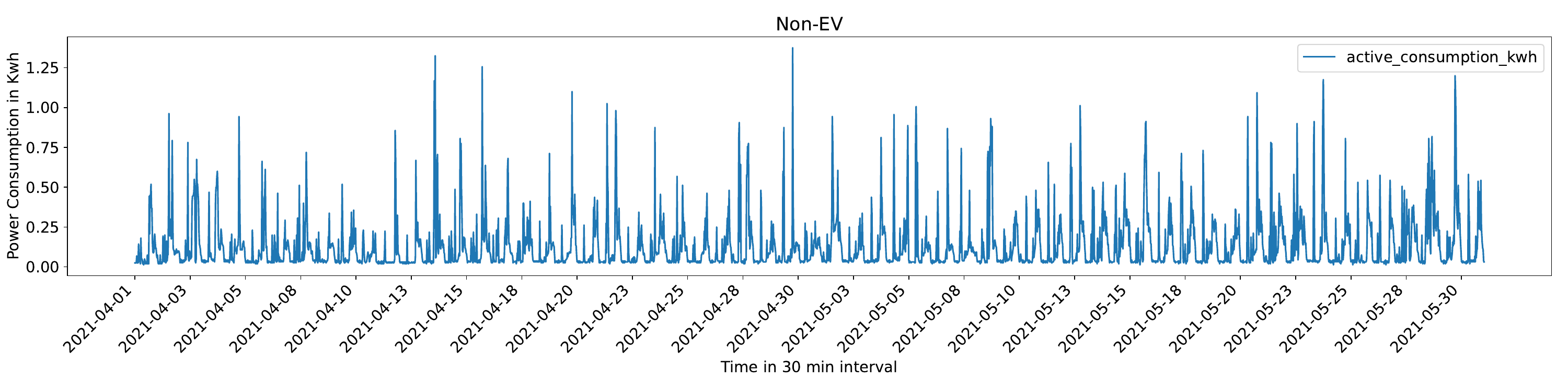}}\\
    \subfloat[Another Non-EV user.]{\includegraphics[width=0.9\textwidth]{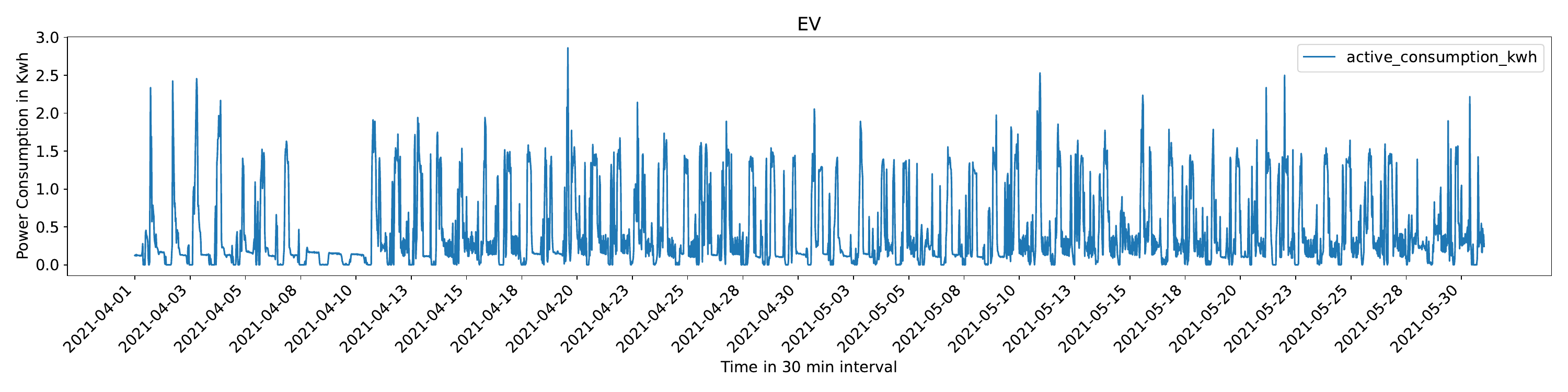}}\\
 	\caption{Power consumption for different households with EV and Non-EV.}
 	\label{EV-User}
 \end{figure*}

In this section, we discuss how the EV charging pattern is different from non-EV-related daily power consumption. EV charging event/activity has the following characteristics:
\begin{itemize}
    \item EV charging is an instantaneous increase in power consumption for a few hours, as shown in~\Fig~\ref{EV-User}.
    \item This instantaneous increase doe not necessarily occurs daily. Also, due to lower electricity prices at night (off-peak hours), some consumers choose to charge their EVs from midnight to early morning. To effectively learn from $X$, we decided to use weekly meter readings and divide the times series $\mathbold{x}_i$ into a set of non-overlapping sequences~$\mathbold{x}_{ij}\subset \mathbold{x}_i$ where $\mathbold{x}_{ij}=\{x_{ij,t},t=1,2,
\ldots, W\}$ where $W=336$ is the number of weekly readings based on 30 minutes reading intervals. The effectiveness of dividing time series into non-overlapping sequences has been shown in the literature~\cite{malhotra2016lstm}. Generally, for each consumer's data $\textbf{x}_i$, we have $M=T/W$ sub-sequences of meter readings.
\item The key challenge here is that different EV users have different power consumption profiles based on EV chargers that have been installed. Some EV users have low-demand consumption, as shown in~\Fig~\ref{EV-User} (b), making it very difficult to identify these users using event-based methods.
\end{itemize}
\subsection{The Proposed Temporal Encoder-Decoder Network}
In the sequence to sequence (seq-2-seq) modeling, we have a sequence of time series defined by $\mathbold{x}_i=(x_1,\ldots,x_{\subseq}) \in \mathcal{X}\subset \mathbb{R}^1 $. For each time step, we want to predict the output sequence as $\hat{\mathbold{y}}_i=(\hat{y_0},\ldots,\hat{y}_{\subseq}) \in \mathcal{Y} \subset \mathbb{R}^1$, in which the output can be a ground-truth label as in a classification task or real value as in a regression task. Formally, we want to find a mapping function $f:\mathbold{x}\to \mathbold{y}$ as follows:
\begin{equation}
    (\hat{y_1},\ldots,\hat{y_{\subseq}})=f(x_o,\ldots,x_{\subseq}),
\end{equation}
where $\subseq$ is the length of the sequences. The learning in seq-2-seq aims to build a network $f(\cdot;\theta)$ parameterised by $\theta$ that minimises a loss function $\mathcal{L}(\cdot)$ between the real output and the predicated output $\mathcal{L}(y_0\ldots,y_{\subseq},\hat{y}_0,\ldots,\hat{y}_{\subseq})$. In the case of auto-encoder, this simplifies into $\mathcal{L}(x_0\ldots,x_{\subseq},\hat{x}_0,\ldots,\hat{x}_{\subseq})$. The above formulation is called causal if it satisfies the casual constrain. This means that to predict $\hat{x}_t$, we use only the previous sequences $(x_0,\ldots,x_{t-1})$. The casual constrain is very important to ensure that there is no leak of future information into the prediction of the current step. Temporal convolution network (TCN)~\cite{bai2018empirical} ensures such a feature compared to using only 1-D convolution along the time axis.

TCN inherits the same properties of convolution neural networks but is mostly used for sequence learning. Generally, TCN comprises a 1-D fully convolution layer (FCN) followed by zero padding to ensure the causal constrain. Here we employ the same TCN block used in ~\cite{bai2018empirical}, which can be described by their elements, the temporal kernel size $k$, a list of a number of filters $n_f=\{n_{f_1},n_{f_2},..\}$, dilation rate $d$ which is determined based on the number of layers $d_i=i*2$ where $i$ is the \textit{i}th layer. 

\begin{figure*}
    \centering
    \includegraphics[width=0.9\textwidth]{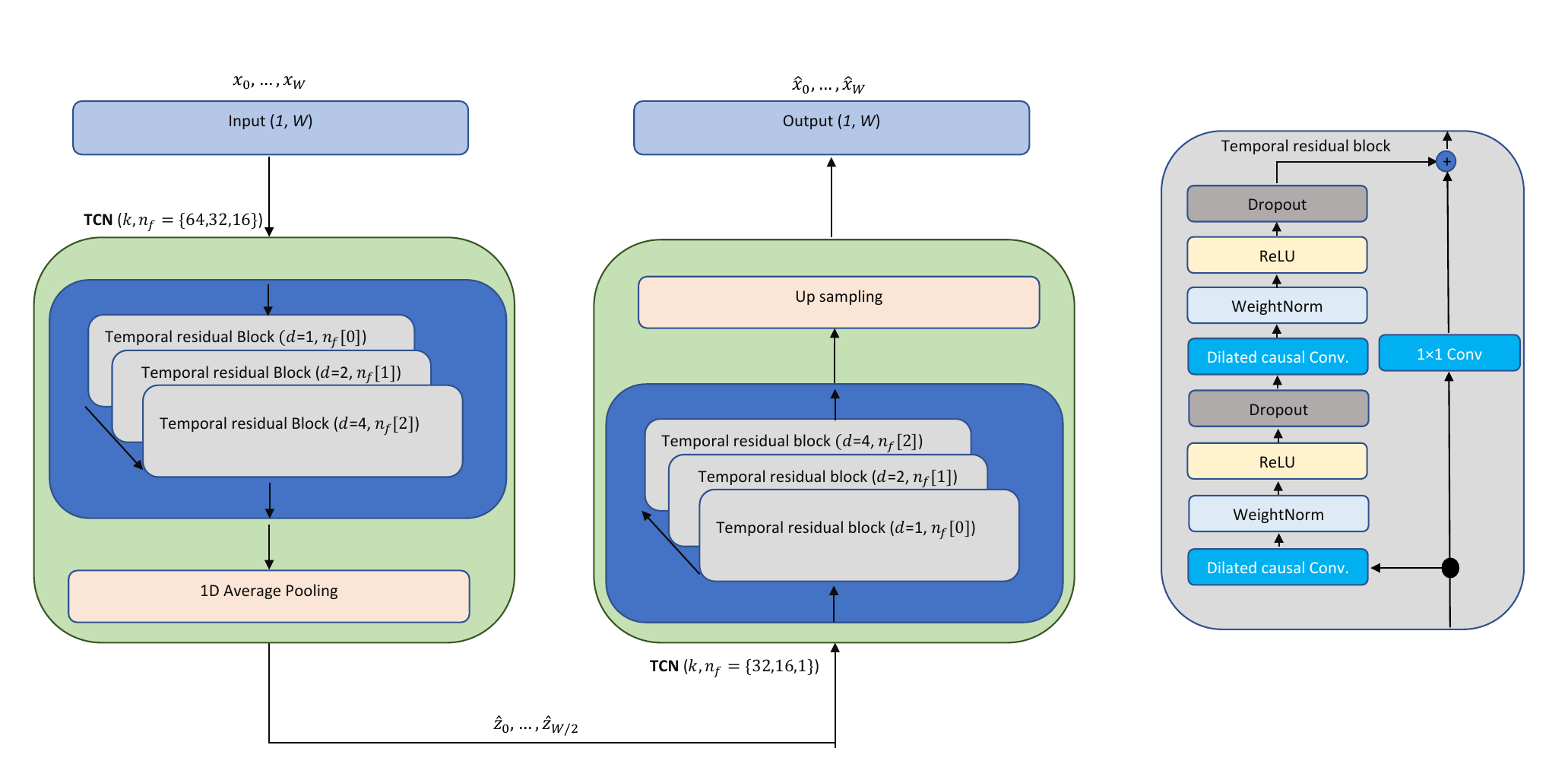}
    \caption{The proposed architecture of TAE.}
    \label{fig:proposed_model}
\end{figure*}

The proposed temporal autoencoder (TAE) architecture is shown in \Fig~\ref{fig:proposed_model}. Typically, the autoencoder consists of an encoder and a decoder. In our proposed network, the encoder and decoder consist of only one TCN block, defined by three temporal residual blocks followed by temporal average pooling/up-sampling. The components of the temporal residual block are dilated casual convolution, weight normalization, ReLU activation\footnote{Rectified linear unit (ReLu)=$f(x) = max(0,x)$}, and finally a dropout layer as shown on the right side of \Fig~\ref{fig:proposed_model}.
Formally, the dilated casual convolution operation $F$ on element $s$ for sequences \mathbold{x} and a filter $C:\{0,...,k-1\}$ is defined as follows~\cite{zhen2019dilated}:
\begin{equation}
    F(s)=\sum_{i=0}^{k-1}C(i) \mathbold{x}(s-id).
\end{equation}
\begin{figure*}
    \centering
    \includegraphics[width=0.9\textwidth]{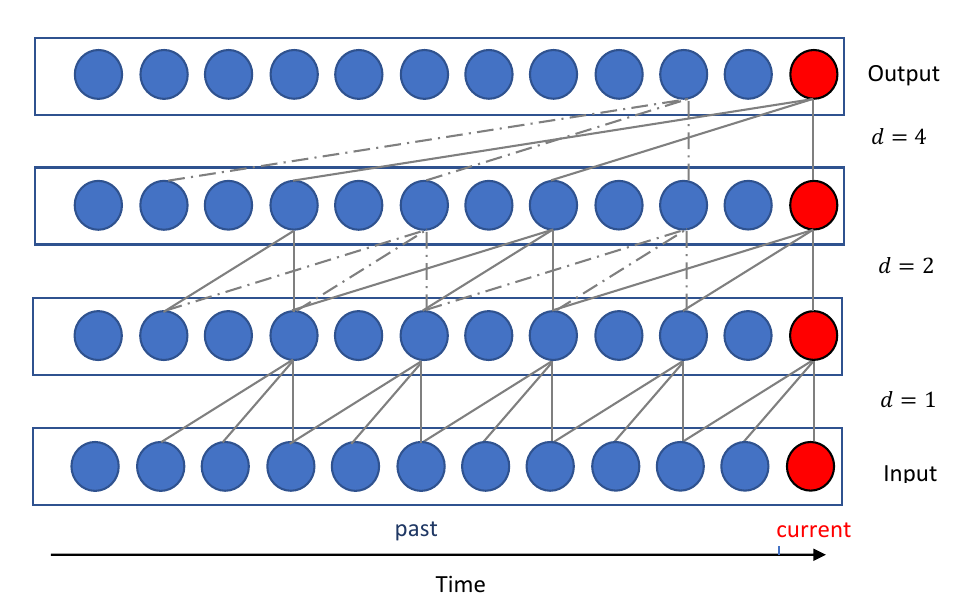}
    \caption{Dilated causal convolution with  different dilation factors $d$ and kernel size of $k=3$.}
    \label{fig:temporal_conv}
\end{figure*}
\Fig~\ref{fig:temporal_conv} illustrates an example of the dilated convolution for different dilation and kernel size $k=3$. The receptive fields can be controlled by increasing the dilation factors, which reflects how much historical information can be used to infer the current time step.
The earlier operation is turned into a normal convolution when a dilation $d=1$. For layer $i$ in our encoder/decoder network, there are a set of  $1$-D temporal filters, parameterized by tensor $W_i\in \mathbb{R}^{n_{f_i}\times k\times n_{f_{i-1}}}$, where $k$ is temporal convolution length and $n_f$ is the number of convolution filters. In our network, we use three residual blocks. The residual connection used in the residual block increases the network stability~\cite{bai2018empirical}.
The output \mathbold{o} of residual block for input sequence \mathbold{x} is expressed as follows:

\begin{equation}
\mathbold{o}=\textbf{max}(0,\mathbold{x}+\mathcal{F}(\mathbold{x})),
\end{equation}
where $\mathcal{F}$ is series of transformations, and \textbf{max} is an activation function. The network weights regularization is controlled by the weight normalization (WeightNorm) and dropout layers.
Generally, the proposed encoder processes a source time sequence with the length of $\subseq$ and outputs a latent/compressed representation of the source as $\mathbold{z}=(\{z_1,\ldots,z_{\subseq}\})$. The decoder takes the latent representation $\mathbold{z}$ and outputs the reconstructed ones $\hat{\mathbold{x}}$.

\subsubsection{Objective function}

The main goal of the loss/objective function is to measure how accurate the model is in making predictions. For the autoencoder, the common loss function is the $\mathcal{L}_2$ norm of the reconstruction loss.~
Given $N$ samples, the reconstruction loss is defined as
\begin{equation}
    \mathcal{L}_{our}(\mathbold{x},\hat{\mathbold{x}})=\frac{1}{N}\sum_{i=1}^{N}||\mathbold{x}_i-\hat{\mathbold{x}}_i||_2.
\end{equation}
We also investigate different combinations of loss functions discussed in~Section~\ref{sec:study_analysis}.

\begin{figure}
    \centering
     \includegraphics[width=0.45\textwidth]{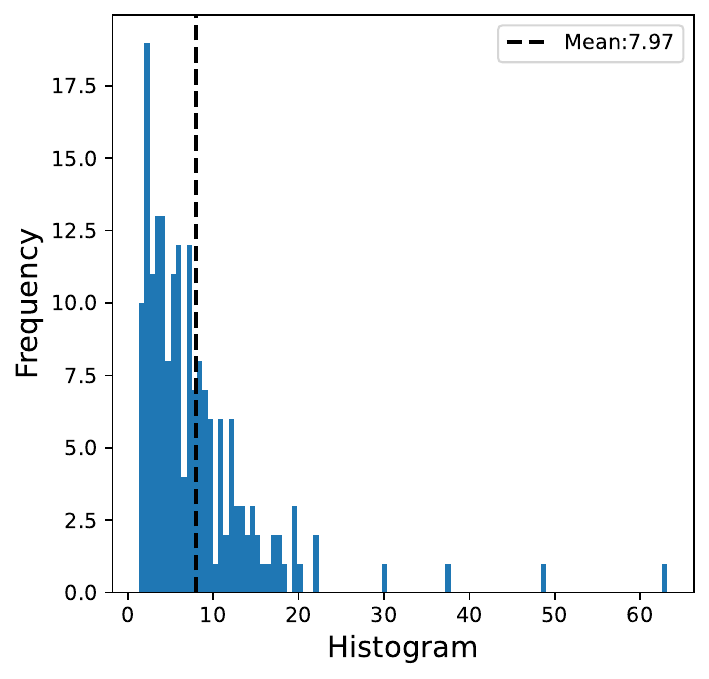}
    \caption{Anomaly score's histogram of the validation set, the mean is used as an anomaly threshold $A_{th}$ in the testing phase.}
    \label{fig:validation_error_hist}
\end{figure}

The general framework for EV user identification using anomaly detection is shown in Algorithm~\ref{alg:1}. We use the proposed user anomaly score $AT$ explained in Section~\ref{sec:anomaly_score} for all models. For our model, the anomaly threshold value ${A}_{\text{th}}$ in Algorithm~\ref{alg:1} is calculated based on the mean of $AT$ for the validation set as shown in~\Fig~\ref{fig:validation_error_hist}.

\begin{algorithm*}[ht]
	\begin{algorithmic}[1]
		\Procedure{Anomaly\_Detection}{$\mathbold{x}_T; \text{TAE}(;\theta) ,{A}_{\text{th}}$}
		\Comment $:\mathbold{x}_T$ power consumption data of length $T$; $\text{TAE}(;\theta):$ TAE network with pretrained parameters, $\theta$; ${A}_{\text{th}}:$ anomaly score threshold.
		\State $X=\{\mathbold{x}_{\subseq 1},\ldots,\mathbold{x}_{\subseq M}\}\gets$ \Call{Process sequence}{$\mathbold{x}_T$}
		\Comment A set of sub-sequences, each of length $\subseq$.
		\State $AT \gets 0$
		\Comment Initialize the anomaly score with zero
 		\For{$\mathbold{x}_{\subseq} \in X$}
 		    \State $\hat{\mathbold{x}}_{\subseq}\gets$ \Call{TAE}{$\mathbold{x}_{\subseq};\theta$}
 		    \Comment Forward pass to generate the reconstructed sub-sequence $\hat{\mathbold{x}}_{\subseq}$.
 		    \State $AS\gets$ \Call{$\mathcal{L}_{our}$}{$\mathbold{x}_{\subseq},\hat{\mathbold{x}}_{\subseq}$}
 		    \Comment Calculate the loss function between the source subsequence and reconstructed ones.
 		    \State $AT \gets AT+AS^2$	
 		    \Comment The anomaly score of $\mathbold{x}_T$.
 		\EndFor
		\If{$AT > {A}_{\text{th}}$} 
		    \State $\eta \gets 1$ 
		    \Comment EV charging detected.
		\EndIf
		\State \textbf{return} $\eta$
		
		\EndProcedure	
	\end{algorithmic}  
	\caption{TAE EV charging identification.\label{alg:1}}
\end{algorithm*}

\section{Experiments}
\subsection{Data pre-processing}
\begin{figure*}
    \centering
     \subfloat[Original smart meter readings and their smoothed version for EV users.]{\includegraphics[width=0.9\textwidth]{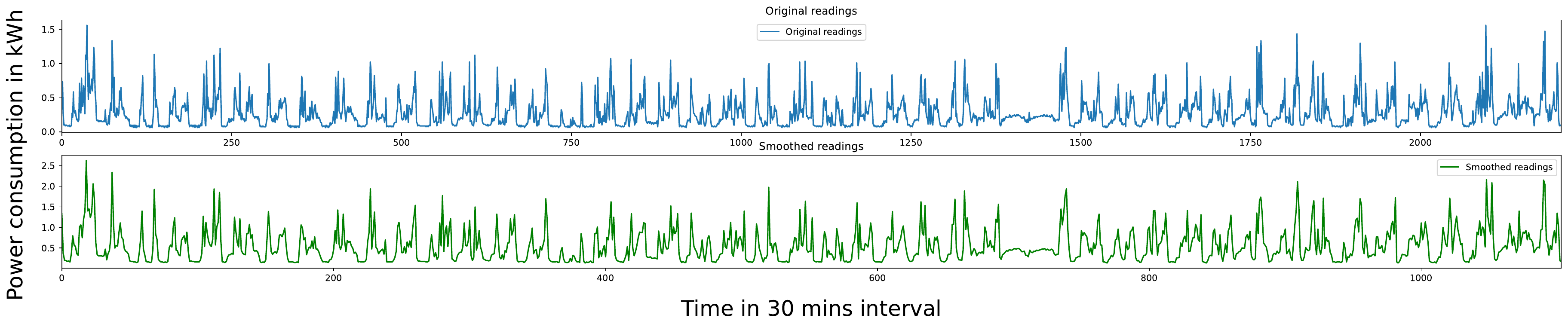}}\\
    \subfloat[Original smart meter readings and their smoothed version for non-EV users.]{\includegraphics[width=0.9\textwidth]{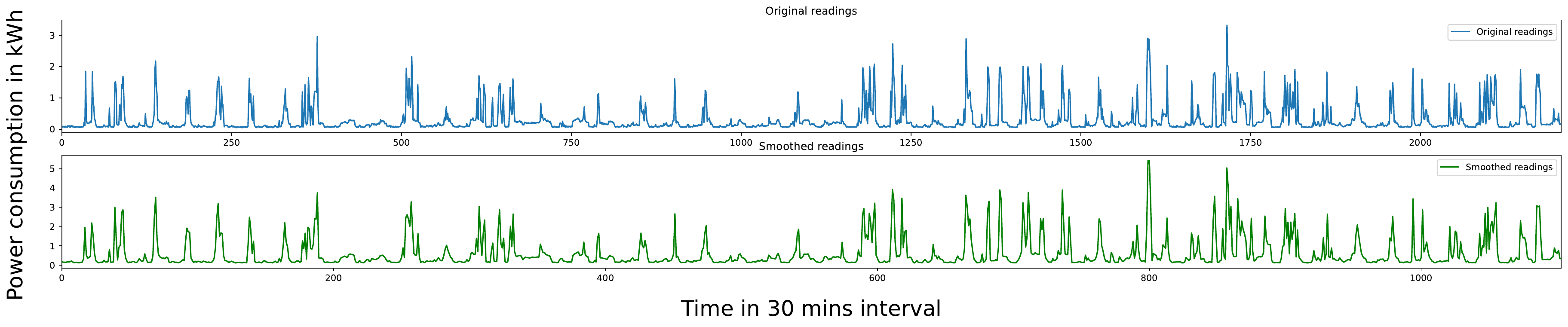}}\\

    \caption{Examples of smart meter readings of EV and non-EV users with their smoothed versions.}

    \label{fig:original_reading_with_their_smoothed_version}
\end{figure*}

Data preparation plays a significant role in the performance of deep learning models. Dealing with our smart meter data, we first perform data smoothing by taking the rolling sum for a window of size two. This step reduces the time series length in half, as shown in \Fig~\ref{fig:original_reading_with_their_smoothed_version}. The intuition behind this step is to remove any steep noise in the data due to an error, or an appliance with high power demand for a short time from the smart meter readings. We then perform data scaling on the entire dataset. Finally, the time series is divided into non-overlapping sequences to be passed into our TAE model.

\label{Sec:experiments}

\subsection{Dataset}
In this paper, we have used net power consumption data from households in Victoria, Australia. The dataset has been collected from 1245 randomly selected consumers between 1 March and 31 May 2021. The smart meter recordings are sampled at 30-minute intervals. The smart meter devices record the power consumption and export from and to the grid. The dataset consists of 1106 non-EV users and 139 EV users. The non-EV users are randomly divided into 774 (70\%) for training and validation and 332 (30\%) users for testing. The test set generally consists of 537 users (332 non-EV and 139 EV users).~Table~\ref{Tab:dataset_stat} shows statistics of testset. The uniqueness of this dataset is the testset has different EV makers and models, some models with high demand, households with wall chargers installed, and others with low demand. 

\begin{table}
\footnotesize
	\caption{Statistics of the EVs' model and their power consumption present in the testset.}
	\label{Tab:dataset_stat}
	\centering
	\begin{tabular}{ccc}
		    \toprule
    		EV Make & EV Model &Power Consumption Type\\
    		\hline
    		\multirow{6}{*}{Tesla}&\multirow{2}{*}{Model 3} & Low demand\\&& High demand\\
    		\cline{3-3}
    		& \multirow{2}{*}{Model X} & Low demand\\&& High demand\\
    		\cline{3-3}
    		& \multirow{2}{*}{Model S} & Low demand\\&& High demand\\
    		\hline
    		\multirow{3}{*}{Hyundia}&\multirow{2}{*}{Kona} & Low demand\\&& High demand\\
    		\cline{3-3}
    		& Loniq& Low demand\\
    		\hline
    		Jaguar& Ipace& High demand\\
    		\hline
            \multirow{2}{*}{MG}&\multirow{2}{*}{ZS} & Low demand\\&& High demand\\
            \hline
            	Nissan& Leaf&Low demand\\
            	\hline
            		Volvo& V60&Low demand\\
            \hline
            Renault&Zoe&High demand\\
            \hline
            Merceedes&GLC300e&Low demand\\
             \hline
            Mini&&Low demand\\
             \hline
            BMW&330e&Low demand\\
             \hline
           Fiat&500e&Low demand\\
           \hline
           Holden& Volt&Low demand\\
           \hline
           Mitsibishi&Outlander PHEV&Low demand\\

		\toprule
	\end{tabular}
\end{table}
\subsection{Anomaly score}
\label{sec:anomaly_score}
This section explains the general framework for identifying an EV user. Given a user with smart meter data of length $T$, there is a total of $M$ subsequences for each consumer. We define the following anomaly score for each user:

\begin{equation}
    AT=\sum_{\mathbf{x}\in M} AS(\mathbold{x}_{\subseq})^2,
    \label{eq:total_anomaly_score}
\end{equation}
where $AT$ is the user anomaly score, which is the squared sum of anomaly scores of each sub-sequence $AS$. The anomaly score of each subsequence is calculated based on $\mathcal{L}_2$. Simply, the $AS$ score is the sum of squared error over the time series, which is written as,
\begin{equation}
    AS=\sum_{\mathbf{x}_i\in \subseq}(||x_i-\hat{x_i}||_2)^2.
    \label{eq:total_anomaly_score_2}
\end{equation}

\subsection{Evaluation protocol}
We consider the common performance metrics for anomaly detection used in the literature~\cite{zhai2016deep,schlegl2019f} to evaluate the proposed  method, which are average precision, recall, and $F_1$ score as metrics for the performance evaluation. We also include the Area Under Receiver Operator Characteristic (AUC)~\cite{huang2005using},  which measures the area under the true positive rate as a function of the false positive rate. The AUC is not sensitive to any threshold or percentage of anomalies present in the test set. The precision and recall are defined as
\begin{equation}
    \text{Precision}=\frac{TP}{TP+FP},  \text{Recall}=\frac{TP}{TP+FN},
\end{equation}
where $TP$ is the true positive rate, $FP$ is the false positive rate, and $FN$ is false negative rate. The $F_1$ is expressed as,
\begin{equation}
    F_1=2\times\frac{\text{Precision}\times \text{Recall}}{\text{Precision}+\text{Recall}}.
\end{equation}

\subsection{Experimental results}
In this section, we explore the performance of the proposed approach against different baseline methods. For a fair comparison, we compare our proposed model with different unsupervised anomaly detection, as listed below.

\begin{itemize}
	\item Local Order Factor outlier detection (LOF): LOF detects anomalies based on the local density deviation of a given point from its neighbours.
	\item Isolated Forest (IF) algorithm: It is based on running several trees, known as Isolated trees.
	\item One-Class Support Vector Machine (OCSVM): OCSVM detects an anomaly when the given data point is far from a decision boundary. 
	\item Long-Short Term Memory autoencoder (LSTM-ED)~\cite{malhotra2015long}: This methodology uses LSTM to encode the time series sequence into latent representation and reconstruct it back. LSTM-ED is trained using Mean Squared Error (MSE) loss function, and the anomaly score is calculated based on the log-likelihood of the test sequence from training error distribution. Long-Short Term Memory autoencoder. LSTM-ED uses LSTM to encode the time series sequence into latent representation and reconstruct it back. LSTM-ED is trained using Mean Squared Error (MSE) loss function, and the anomaly score is calculated based on the log-likelihood of the test sequence from training error distribution.
	\item Deep autoencoder with Gaussian Mixtures Models (DaGMM)~\cite{zong2018deep}: The distribution assumption of these models has been imposed on the latent space. DaGMM uses a combination of the reconstruction loss, sample energy, and GMM covariance to train the model. In DaGMM, the sample energy is used as an anomaly score. We have used the Pytorch implementation of the model on GitHub\footnote{https://github.com/mperezcarrasco/PyTorch-DAGMM}. Table~\ref{tab:baseline_settings} shows the parameter settings of baseline models.
\end{itemize}
 For LOF, IF, and OCSVM, we have used sklearn\footnote{https://scikit-learn.org/} library implementation and investigated different parameter settings. Parameter values provided in Table~\ref{tab:baseline_settings} are the optimum ones that result in the best accuracy.

\begin{figure}
    \centering
     \includegraphics[width=0.45\textwidth]{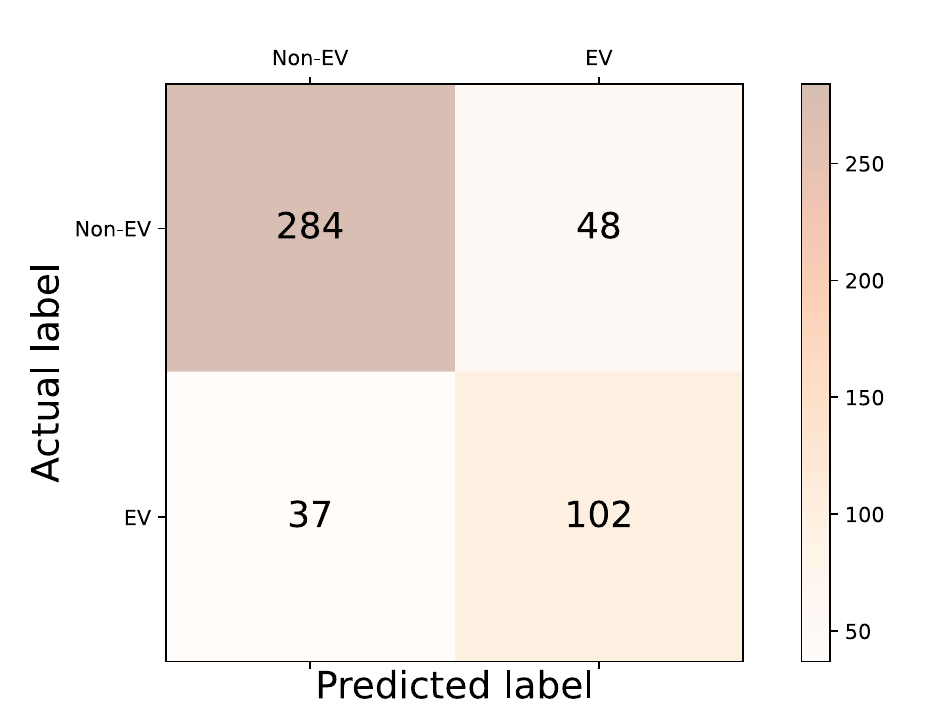}
    \caption{Confusion matrix of the proposed model.}
    \label{fig:confusion_matrix}
\end{figure}

\begin{figure*}
    \centering
     \subfloat[True Negative example (non-EV detected as non-EV).]{\includegraphics[width=0.9\textwidth]{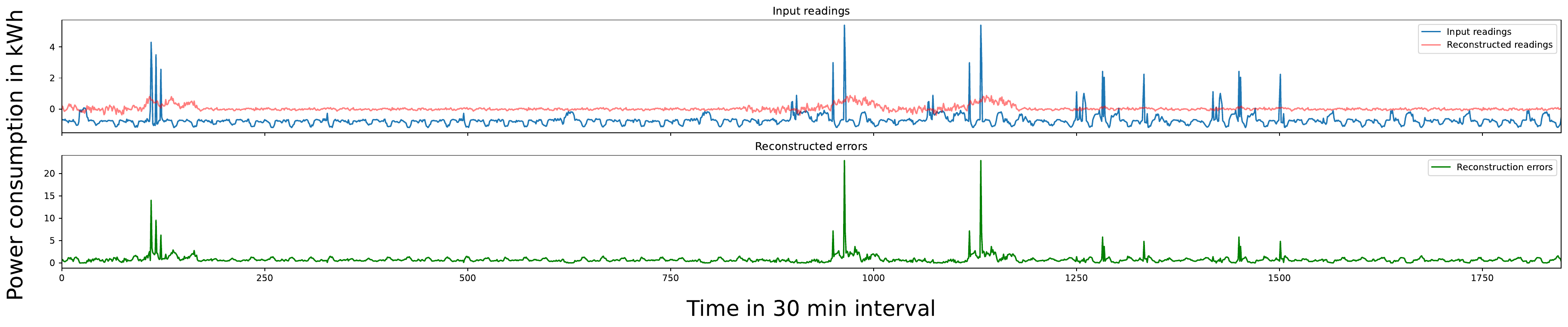}}\\
     \subfloat[False positive example (non-EV detected as EV).]{\includegraphics[width=0.9\textwidth]{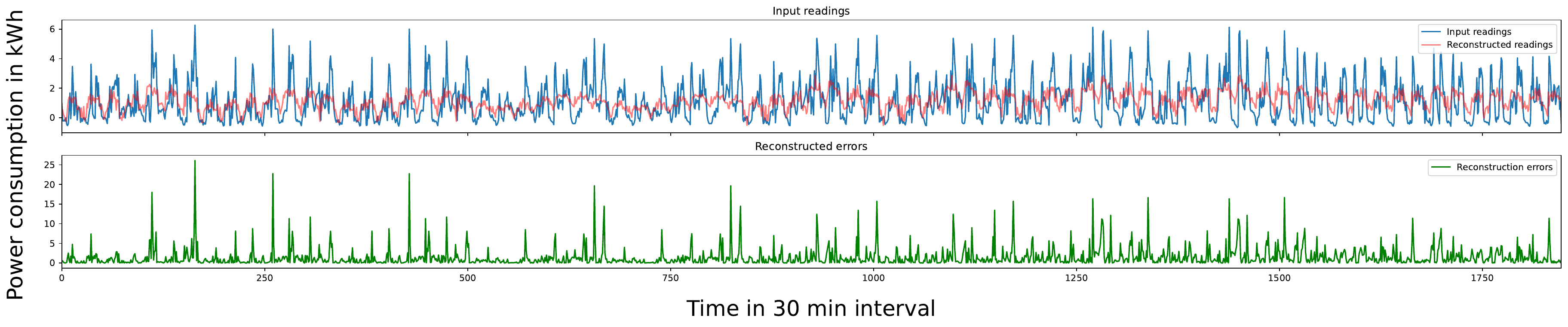}}\\
      \subfloat[False negative example (EV detected as non-EV).]{\includegraphics[width=0.9\textwidth]{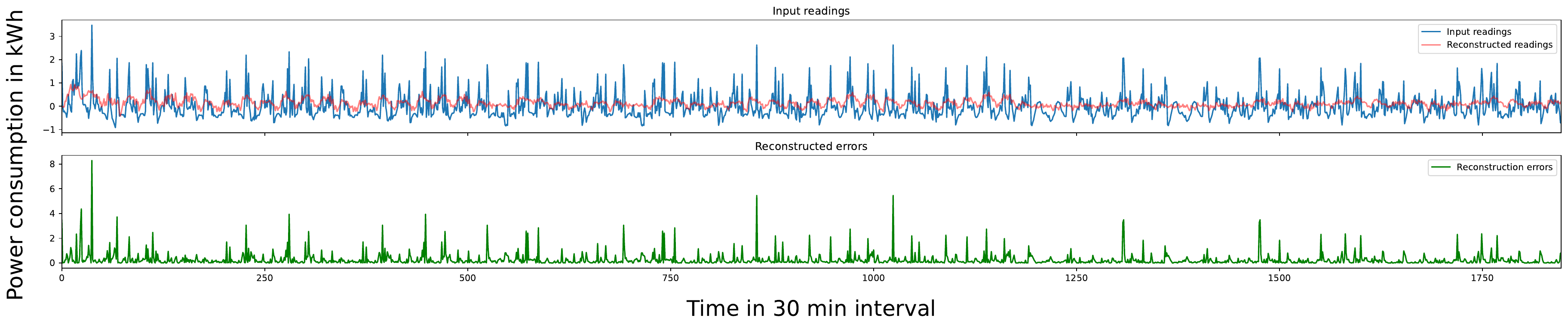}}\\
    \subfloat[True positive (EV detected as EV).]{\includegraphics[width=0.9\textwidth]{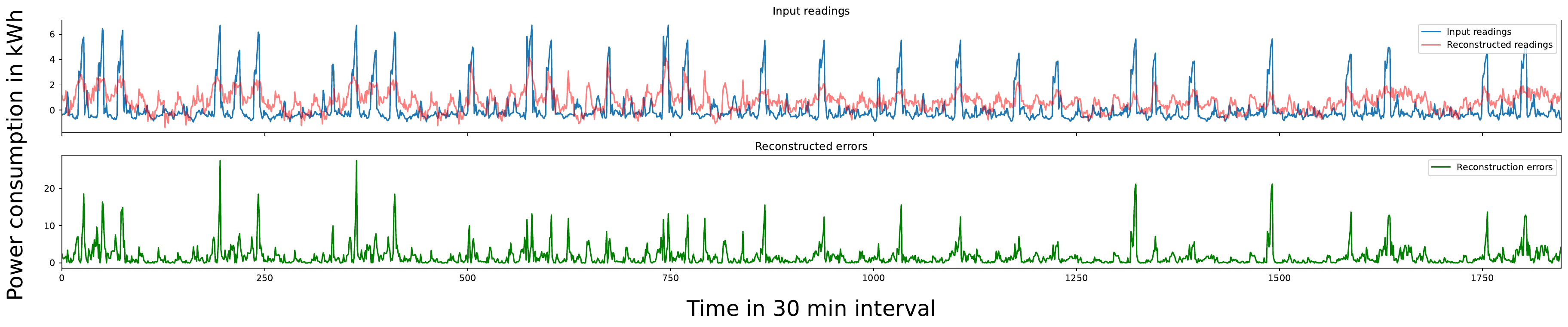}}\\
    \caption{Examples of true negative, false positive, false negative, and true positive with their input, reconstructed readings, and reconstruction errors of our model.}

    \label{fig:fp_fn}
\end{figure*}

Table~\ref{Tab:Powercore_results} shows the performance of our proposed TAE model against different baseline methods, where the best values are shown in bold. TAE has the highest score considering Precision, $F_1$, and AUC metrics. The second-place model is LSTM-ED with $66.67\%$ and $82.47\%$ for $F_1$ and AUC, respectively. The lowest performance model is for LOF. The confusion matrix of the proposed model is shown in \Fig~\ref{fig:confusion_matrix}. The top left of the confusion matrix shows the true negative  (non-EV users correctly classified), the right-top is the false positive (non-EV users misclassified as EV users), the bottom left is the true positive (EV users correctly classified), and the bottom right is the false negative (EV users misclassified as non-EVs). From~\Fig~\ref{fig:confusion_matrix}, our model obtain a true positive = 102, true negative = 284, false positive = 48, and false negative = 37. \Fig~\ref{fig:fp_fn} shows examples of true negative, false positive, and false negative generated by our model. \Fig~\ref{fig:fp_fn}~(a) shows the power consumption of a non-EV user, which is successfully identified using our model. This consumer has smooth power consumption with few peaks in consumption, and the model is able to reconstruct the signal. \Fig~\ref{fig:fp_fn}~(b) shows a false positive example where the input readings have a lot of fluctuations in consumption, and the model is unable to reconstruct and thus generates high reconstruction errors. Another mis-detection is shown in~\Fig~\ref{fig:fp_fn}~(c), where the model again fails to reconstruct the input signal and generates a low reconstruction signal.

\begin{figure}
    \centering
    \includegraphics[width=0.45\textwidth]{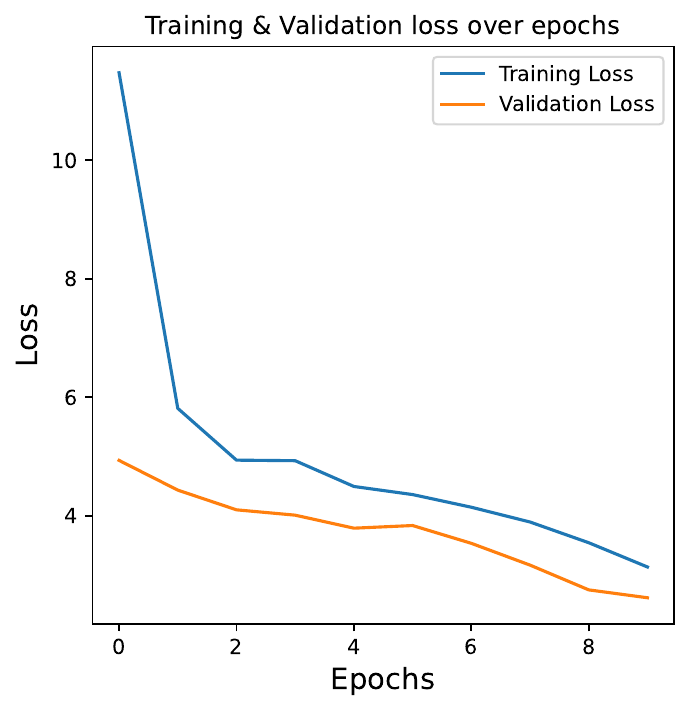}
    \caption{Training loss vs validation loss of our model.}
    \label{fig:train_valid_loss}
\end{figure}

Generally, it can be seen that our model has decent precision and recall. \Fig~\ref{fig:train_valid_loss} shows the training loss and validation loss in which the model is not overfitting during the training phase.

\begin{table}
\footnotesize
	\caption{Precision, Recall, $F_1$, and AUC in \% for detected anomalies on the test set, the best results are in bold.}
	\label{Tab:Powercore_results}
	\centering
	\begin{tabular}{ccccc}
		    \toprule
    		Methods & Precision \% & Recall \% &$F_1$ \%& AUC \% \\
    	\toprule
    	LOF & 50 & 1.44 & 2.80 & 28.11 \\
    	IF& 29.42&99.28&45.39&29.20\\
		OCSVM~& 33.33&7.19 &11.83 &32.75\\
		LSTM-ED~\cite{malhotra2016lstm}& 62.66 & 71.22 & 66.67& 82.47\\
		DaGMM~\cite{zong2018deep}&46.50& 66.91 & 54.87&70.61\\
		1D-CNN-AE&48.73&69.06&57.14&69.43\\
		TCN~\cite{thill2020time}&53.11  &  67.63&59.49&76.04\\
		\hline
		$\mathcal{L}_{our}$&\textbf{69.66}&72.66& \textbf{71.13}& \textbf{85.10} \\
		\toprule
	\end{tabular}
\end{table}

\begin{table*}
    \centering
    \footnotesize
    \caption{Parameter settings for baseline models.}
    \begin{tabular}{cc}
    \toprule
       Methods  & Settings\\
      \toprule
         LOF& Number of neighbors:10, novelty:True\\
         IF& Number of estimators:100, contamination:0.01\\
         OCSVM& Kernel:"rbf", gamma:0.6\\
         LSTM-ED& Number of LTSM layer: 2, hidden size: 5,\\
         & solver={Adam}, learning rate (lr):1e-3\\
         DaGMM& Number of GMM= 2, MLP layer size=\{168,60,30,10,1\}, \\
        & solver={Adam}, learning rate (lr):1e-3\\
         TCN &Loss=\{MSE\}, Number of TCN blocks:1,\\
         &kernel size:20, dilatation list:\{1,2,4,8,16\}\\
         \toprule
    \end{tabular}
    \label{tab:baseline_settings}
\end{table*}
\subsection{Study analysis}
\label{sec:study_analysis}
We further investigate the use of different loss functions for training the model and also use these loss functions in the test phase to calculate the anomaly score. The $L_2$ loss function is a Euclidean distance measure that has its own disadvantages, such as sensitivity to scaling and shift-invariance. In contrast, Dynamic Time Warping (DTW) distance is a similarity distance that has been shown to be very successful for time series pattern mining and measuring the similarity between time series~\cite{maghoumi2021deepnag}. DTW is known to be scale- and time-shift invariant. It is computed via Dynamic Programming (DP), which involves calculating a cost function $\Delta (\cdot,\cdot)$ by computing $n\times m$ pairwise distances, where $n$ and $m$ are the lengths of the source and target signals. The standard DTW calculation is computationally expensive, and most importantly, is not differentiable to be used as a loss function for deep learning models. Cuturi~\andothers~\cite{cuturi2017soft} proposed a soft-DTW (SDTW) as an efficient and differentiable version of DTW to be used as a loss function. The SDTW loss function is defined as follows: 
\begin{equation}
    \mathcal{L}_{SDTW}(\mathbold{x},\hat{\mathbold{x}})=min^\gamma\{\langle A,\Delta(\mathbold{x},\hat{\mathbold{x}})\rangle,\mathcal{A}\in \mathcal{A}_{n,m}\},
\end{equation}
where $\gamma >= 0$ is the smoothing parameter, $A \in \mathcal{A}_{m,n}$ is an alignment matrix, $\mathcal{A}_{m,n} \subset \{0,1\}^{n \times m}$ is the set of binary alignment matrices. For more information on how to calculate the forward and backward pass of $\mathcal{L}_{DTW}$ see~\cite{cuturi2017soft}. 


Besides the soft-DTW, we investigate different combinations of the loss on the accuracy of the model. The final loss function $\mathcal{L}_{final}$ is expressed as,
\begin{equation}
    \mathcal{L}_{final}=\lambda_1\mathcal{L}_{rec}+\lambda_2\mathcal{L}_{DTW}++\lambda_3 \mathcal{L}_{cos},
\end{equation}
where $\lambda_1,\lambda_2,$ and $\lambda_3$ are hyperparameters that control the contribution of each particular term into the final loss function. $\mathcal{L}_{cos}$ is the cosine similarity loss function defined as,
\begin{equation}
\mathcal{L}_{cos}(\mathbold{x},\hat{\mathbold{x}})=\frac{1}{N}\sum_{i=1}^{N}\frac{\mathbold{x}\cdot \hat{\mathbold{x}}}{||\mathbold{x}||_2 ||\hat{\mathbold{x}}||_2},
\end{equation}
where $\cdot$ is the element-wise multiplication. 

Table~\ref{Tab:different_loss_funs} shows the performance of our model (TAE) using different metrics with different loss functions. The top part of Table~\ref{Tab:different_loss_funs} shows the performance using only one loss function; TAE $(\mathcal{L}_{our})$ has the best performance in all metrics. We also recorded the total training run-time using HP EliteBook X360 with 8G RAM and Intel Core-i5-$8^{th}$ generation. The total run-time for all models for 10 epochs is shown in the last column. Our loss function is the most computationally efficient compared to others. We observe that training with $\mathcal{L}_{DTW}$ is computationally expensive and challenging due to the fact that $\mathcal{L}_{DTW}$ can be sometimes negative.

\begin{table*}
\footnotesize
	\caption{Performance of different loss functions on our TAE model.}
	\label{Tab:different_loss_funs}
	\centering
	\begin{tabular}{cccccc}
		    \toprule
    		Loss function & Precision \% & Recall \% &$F_1$ \%& AUC \%&Training time in sec.\\
    	\toprule
	    $\mathcal{L}_{cos}$& 30.43& 5.04 &8.64  &42.80&82.63\\

		$\mathcal{L}_{DTW}$&77.78  &30.22&43.52 &52.75  &977.76\\
		$\mathcal{L}_{our}$& 69.66&72.66& \textbf{71.13}& \textbf{85.10} &69.60\\
		\hline
 		 $\mathcal{L}_{final}$&  &  &  & &\\
 		 $\lambda_1=1$,$\lambda_2=0$,$\lambda_3=1$ &66.67  &70.50  & 68.53 &84.26 &78.51\\
 		 $\lambda_1=1$,$\lambda_2=1$,$\lambda_3=0$& 87.80 & 25.90 & 40.00&42.78&1013.01\\
 		 $\lambda_1=0$,$\lambda_2=1$,$\lambda_3=1$& 77.78 & 30.22 & 43.52&52.75&1354.51\\
 		 $\lambda_1=1$,$\lambda_2=1$,$\lambda_3=1$& 80.77&30.2 &43.98 &48.50&1101.37\\
		 \toprule
\end{tabular}
\end{table*}

\section{Conclusion}
\label{Sec:conclusion}
This paper proposed an unsupervised anomaly detection technique for EV identification from behind smart meter data. Considering that EV charging may happen at any time of the week, we divided the smart meter data into non-overlapping sequences, where each sequence represented readings over one week. We proposed a temporal autoencoding model using a temporal convolution neural network to compress the non-EVs smart meter data into latent representations. We proposed an $L_2$-based loss function to train the model. EV charging was identified as an anomaly when the model returns a high anomaly score. The Anomaly score was calculated by taking the squared sum of the reconstructed error over the entire meter readings. The proposed model was benchmarked against several unsupervised learning models and showed superior performance in terms of $F_1$ and AUC metrics. Also, we showed that for this particular problem, the proposed loss function was more efficient compared to Soft-DTW.

Potential future works are i) building anomaly score that does not take the contribution of reconstructions errors equally into the final ones, by utilizing other than the mean operation such as using attention mechanism, ii) Studying the effect of uncertainties, such as solar generation, on the overall performance of the anomaly detection model and iii) developing an EV detection model using daily instead of weekly measurements.


%





\bibliographystyle{IEEEtran}
\bibliography{references}

\newpage
\end{document}